\documentclass[manuscript]{aastex}
\slugcomment{Submitted to the Astronomical Journal}
\received{2012 August 12}
\accepted{2012 December 11}
\shorttitle{Remote Globular Clusters in M31}
\shortauthors{di Tullio Zinn \& Zinn}
\begin{document}
\title{MORE REMOTE GLOBULAR CLUSTERS IN THE OUTER HALO OF M31}
\author{Graziella di Tullio Zinn and Robert Zinn}
\affil{Department of Astronomy, Yale University, P.O. Box 208101, New Haven, CT 06520-8101}
\email{graziella.zinn@yale.edu}
\begin{abstract}
We searched the Sloan Digital Sky Survey for outer halo globular
clusters (GCs) around M31.  Our search of non-stellar objects, within
the limits of $0.3\le (g-i)_{0} \le 1.5$ and $14.0\le r_{0} \le 19.0$
concentrated in some remote areas of the extended halo, to a maximum
projected distance of 240 kpc, for a total of approximately $200\, \textrm{deg}^2$.  Another $\sim 50\, \textrm{deg}^2$, $\sim5-75$ kpc from M31, were
surveyed as test areas.  In these areas, we identified 39 GCs and 2 GC
candidates $84\%$ of the previously known GCs (93\% of the``classical
GCs'' and 40\% of the ``halo extended clusters'', on the
cluster classification scheme of Huxor et al.).  For the entire survey, we
visually inspected 78,516 objects for morphological evidence of
cluster status, and we identified 18 new clusters, and 75 candidate
clusters.  The new clusters include 15 classical globulars and three
clusters of lower density.  Six of the clusters reside in the remote
areas of the outer halo, beyond projected distances of 100 kpc.
Previously, only MGC1 was found beyond this limit at 117 kpc.  The
farthest cluster discovered in this survey lies at a projected radius
of 158 kpc from M31, assuming that the M31 distance is 780 kpc.
\end{abstract}


\section{Introduction}

Globular clusters (GCs) are some of the most beautiful objects and are
important probes of galaxy evolution.  In general, GCs are luminous
and compact objects that can be easily identified in nearby galaxies.
The stellar population within a typical Milky Way (MW) GC is simple in
the sense that it is describable by one age and one chemical
composition.  Consequently, the abundances of Fe and the other heavy
elements in a GC probably indicate the composition of the gas in its
host galaxy at the time when the cluster formed. With this
assumption, many studies have used measurements of GCs as diagnostics
of the ages, metallicites, and kinematics of the stellar populations
in Local Group galaxies and in more distant elliptical and spiral
galaxies, including ones in the closest galaxy clusters (e.g.,
Caldwell et al. 2011; see Brodie \& Strader 2006 for a review).

For testing the properties of GCs as tracers of the overall evolution
and assembly of galaxies, the best galaxy is proving to be M31, the
nearest large galaxy to the MW, hosting the most extensive
GC system in the Local Group.  In recent years, several
searches have mapped many new features of the halo of M31 up to an
average projected radius of $\sim150$ kpc.  For example, the Pan
Andromeda Archeological Surveys, conducted with the Mega
Prime (Mega Cam wide-field camera on the Canada-France-Hawaii
Telescope), have discovered 40 new GCs in the outer halo
of M31 \citep{hux08}; multiple tidal debris streams spatially
associated with GCs \citep{mac10}, and five new dwarf satellite
galaxies, Andromeda XXIII-XXVII \citep{rich11}.  The \emph{Hubble Space
Telescope (HST)} Advance Camera for Surveys imaging has confirmed
the presence of a new population of extended, diffuse GCs in M31
larger for their luminosity than any previously known GCs
\citep{tan12}.  M31 is also remarkable in that it contains many GCs in
its remote halo that resemble in luminosity and structure the
``classical GCs'' that populate the inner halos of M31 and the MW
\citep{hux08,hux11}.  In contrast, the remote GCs in the MW are, with
only the exception of giant GC NGC 2419, much sparser and lower in
luminosity than the typical classical GC \citep{hux11}.

In this paper we present the results of a new search for GCs, extended
to more remote regions of the M31 outer halo, up to a maximum distance
of 240 kpc.  Table 1 lists some basic properties of the newly
discovered clusters.

\section{The Search for M31 Remote Clusters}

Our search is based on the Sloan Digital Sky Survey (SDSS) images in
Data Release 8 (DR8; \citet{aih11}).  The most important properties of
the SDSS for our study are its large sky coverage, its near uniform
resolution ($0\farcs40 \times 0\farcs40$ pixels with a median seeing of
$1\farcs40$ in the r-band), and its depth (r=22.2, 95\% completeness
limit Adelman-McCarthy et al. 2006).  Unfortunately, the SDSS
footprint does not cover uniformly the sky around M31, which limited
the extent of our survey.

We selected our sample from the objects classified as non-stellar by
SDSS and therefore included in their galaxy catalogue.  GCs are
expected to be distributed over specific ranges of absolute magnitude
($-10.5<M_{V}<-3.5$,\citet{hux08}) and color.  For placing
corresponding limits on the apparent magnitudes of the M31 clusters,
we chose the r-band because GCs are red objects and because the r-band
has the best signal to noise ratio of the longer wavelength SDSS
bands.  As an indicator of GC color, we preferred g-i, because it has
a long baseline with good signal-to-noise ratio.  By combining the
$(g-r)_{0}$ and $(r-i)_{0}$ colors in Figure 1 of \citet{pea11}, it is
seen that $0.3\le (g-i)_{0} \le 1.5$ encompasses the full range of the
M31 GCs.  We opted to use the Petrosian magnitudes provided by SDSS
because they (1) include most of the light of an extended source, (2)
have high signal-to-noise ratios over the magnitude range, and (3)
measure the flux within the same aperture size for all filters.

At the beginning of our search for GCs we queried the SDSS galaxies
catalog for extended sources, using the complete expected range of
apparent magnitudes for GCs, $14 < r_{0} < 20.5$, assuming a distance
of 780 kpc for M31 \citep{mcc05}.  After some testing with the
confirmed M31 GCs to gain experience with their appearance in the SDSS
images, we realized that we were unable to classify routinely as
clusters objects fainter than $r_{0} = 19.0$, because it was
impossible to distinguish them from background low-surface brightness
galaxies, as in the case of HEC2 \citep{hux08} or from the unresolved
compact galaxies, as in the case of B531 (the Revised Bologna
Catalog v5, RBCv5; Galleti et al. 2004).  We then fixed the range of our
cuts at $14.0\le r_{0}\le 19.0$ and $0.3\le (g-i)_{0} \le 1.5$, aware
that we were imposing another limitation on the completeness of our
sample.  At this time, we did not use ellipticity as a criterion, in
order not to eliminate possible GCs that could appear as somewhat
elliptical objects because of overlapping images of stars.  Even with
our cuts, the large area of our survey area ($250\, deg^2$) left us
with the great majority of the objects in our sample as background
contaminants, mostly faint or compact unresolved galaxies, but also
some large images of stars due to poor seeing, which were detected as
extended sources by SDSS.

We had to visually inspect for morphological evidence of cluster
status a sample of 78,516 non-stellar objects according to SDSS.  This
was accomplished in a series of steps.  The coordinates of the objects
that met our magnitude and color cuts were loaded into the DR8 Image
List tool on the SDSS website.  The image cutouts that were returned
were then scanned by eye, which rejected the vast majority of objects
as not candidate clusters.  The cutouts of the objects that were not
rejected were studied is some detail, and comparisons were made with
the cutouts of known M31 clusters.  For the objects that passed this
more detailed scrutiny, we downloaded the r-band fits images from the
SDSS Web site, and in some cases also the i-band images.  These were
scrutinized in great detail by setting the scaling and zoom to
different levels and by comparison with similar images of known M31
clusters.

As noted by previous surveys \citep[e.g.,][]{batt80}, the appearances
of the M31 GCs vary with their magnitudes and central concentrations,
and also with the plate scale, seeing, and magnitude depth of the
images.  The following classifications were developed after examining
the image cutouts and the r-band fits images from the SDSS of the M31
GCs in the lists of \citet{hux08}, with a few additional GCs from the
RBCv5. (1) GCs: objects clearly resolved into stars with
a concentration in the center or presenting a core with uneven
contours and surrounded by a few point sources, that were identified
as stars by SDSS photometry. (2) High confidence (HC) Candidate
clusters: objects with uneven contours, but less so than category (1),
with or without surrounding stars, but still presenting an overall
appearance of cluster status when the contrast and scaling were
changed. (3) Candidate clusters: objects with a compact shape with no
sign of galaxy structure and not associated with galaxies, or objects
displaying a diffuse nature, with a presence of point sources that
could be stars.  We found it more difficult to classify clusters of
low central surface brightness.  They presented a diffuse appearance,
similar to low surface brightness galaxies, unless they showed some
resolution into stars.  In cases of diffuse objects, we often relied
on comparison with the images provided by \citet{hux08} for the
diffuse and extended clusters that they found, which they labeled
``halo extended clusters'' (HEC), and the SDSS images of these
clusters.  For our classification we prefer to use only the term
``diffuse'', without ``extended'', because the half-light radii,
$R_{h}$, that we measured for our clusters of this type (8.6-10.5 pc)
are smaller than the ones measured by \citet{tan12} for the extended
clusters ($ >18$ pc).  We are not sure if this is an intrinsic
difference of the clusters, or if it is due to the limitation of our
measurements (see Section 3).

Figure 1 shows in rectangular coordinates relative to M31 as the
origin, the distribution of non-stellar objects that passed our
magnitude and color cuts in the survey areas covered by SDSS.  These
encompass a total area of $250\, \textrm{deg}^2$.  The central and middle areas
within $9\degr < RA < 20\degr (-1\degr < \xi < 8\degr)$ and $35\degr <
Dec <45\degr (-5\degr < \eta <4 \degr)$, which span $\sim5-75$ kpc in
projected distance from the center of M31, were selected to test the
efficacy of our search criteria for GCs.

These test areas contain 51 objects that are listed as confirmed GCs
in the RBCv5 ($f=1$ and $c=1$ or $f=8$ and $c=8$).  Forty-six are
``classical GCs'' ($f=1$) in the terminology of \citet{hux08}, and 5
are classified as ``extended clusters'' ($f=8$).  Classical GCs have
relatively bright central surface brightnesses and typically $R_{h}
\lesssim 8$ pc, whereas extended clusters have very faint central
surface brightnesses ($>22 \mathrm{mag\, arcsec^{-2}}$) and $R_{h} >
18$ \citep{tan12}.  Forty-four or $96\%$ of the classical GCs and 2 or
$40\%$ of the extended clusters are in the SDSS galaxy catalogue and
passed our color and magnitude cuts.  The 2 missing classical GCs were
not identified as objects by the SDSS.  They lie relatively close to
M31 where the stellar density is very high.  Since our survey fields
have much lower densities than these fields, we will assume in the
following that the SDSS is complete to our magnitude limit.  Only two of
the five extended clusters in the test areas are brighter than our
magnitude limit.  Both were identified as star clusters by our
examination of the SDSS images, which suggests that our efficacy for
identifying extended clusters is $\sim40\%$.  The visual examination
of the SDSS images of the 44 classical GCs contained in our test areas
yielded 39 definite clusters ($88.6\%$), two candidate clusters
($4.5\%$), and rejected three as probably galaxies ($6.8\%$).  H17
\citep[see image in][]{hux08} is one of the three rejected clusters, which
we could not distinguish from compact galaxies.  If the assignment of
at least cluster candidate status is considered success, then these
results suggest that the efficacy of our search techniques is
$\gtrsim90\%$ for classical GCs and $\sim83\%$ for clusters of all
types.  Since this seemed adequate, we proceeded to search areas of
the remote halo of M31.

The vast majority of the objects that we examined were rejected
because they did not meet our criteria for either GCs or cluster
candidates (see above).  After cross-checking with the RBCv5 and
removing previously identified objects, we were left with 18 new
clusters and 75 new candidate clusters, of which 23 are classified as
High Confidence Candidates.

\section{Results and Discussion}

The new 18 clusters discovered in our search are listed in Table 1; 15
are classical GCs and three are diffuse.  Also listed are the foreground
extinction corrected r and g-i, obtained from the Petrosian magnitudes
and the interstellar extinctions from \citet{sch97} that are listed by
SDSS for each object.  The Petrosian magnitudes are expected to
include about 90\% of the cluster light.  The extinctions, which are
from dust maps, indicate that the MW extinction is modest in the
directions of the clusters ($0.13\leq A_{r} \leq 0.35$).  We caution,
however, that these maps do not resolve the small-scale variations in
extinction that are found in the remote halo of
M31\citep[see][]{mack09}, and it is possible that the clusters with
the smallest projected galactocentric distances, $R_{gc}$, may suffer
some additional extinction from dust in M31.  Also listed in Table 1
are the values of $R_{gc}$, $R_{h}$, and the r band absolute
magnitudes, $M_{r}$, which were calculated assuming a distance of 780
kpc \citep{mcc05}.

The half-light radii are our estimates from the radial profiles that
SDSS provides for each object in their PhotoProfile catalog.  This
catalog tabulates the azimuthally averaged mean flux in concentric
annuli \citep[see][]{sto02}, which we used to calculate the cumulative
distribution of the flux with radius and the average surface
brightness in magnitudes $\textrm{arcsec}^{-2}$. in the annuli.  We fitted a
cubic spline to the cumulative distribution and adopted the radius that
contains 0.5 of the total flux as $R_{h}$.  We also examined a plot of
surface brightness against radius for each object, which indicated
that the $R_{h}$ values for SDSS2 and SDSS7 are more uncertain than
the others.  To check our procedures, we determined values of $R_{h}$
for the eight GCs with SDSS data that were included in the study by
\citet{tan12}, who measured $R_{h}$ from \emph{HST} observations.  Although
the SDSS profiles are shallow ($\lesssim 27 \mathrm{mag\,
  arcsec^{-2}}$) in comparison to the \emph{HST} ones ($\lesssim 30
\mathrm{mag\, arcsec^{-2}}$), they yielded estimates of $R_{h}$
that were within 15\% of the ones determined empirically by
\citet{tan12} for the six GCs with $R_{h} < 10$ pc.  The $R_{h}$ values
that we obtained for the remaining two, which are large extended
clusters, are about 40\% smaller than the empirical measurements by
\citet{tan12} and about 20\% smaller than their values from King model
fits.  Since our sample of clusters does not contain such large, low
surface brightness objects, the $R_{h}$ values in Table 1 should
provide a rough measure of the sizes of the clusters.
                                                                                
The new clusters appear to be below average to average in luminosity,
except for SDSS11, a classical GC, which exhibits an absolute r band
magnitude of -8.9, a half-light radius of 4.0 pc and a projected
galactocentric distance of 44 kpc from M31.  The half-light radii of
the clusters span from 4.0 pc to 10.5 pc.  The diffuse clusters are
among the most extended of the group, but are less extended than the ones
studied by \citet{tan12}, which could be a result of the limits
imposed by our selection criteria.  Our search routine recovered the
brightest clusters of the \citet{hux08} sample of extended clusters
within the SDSS footprint, but it missed the faintest ones.  These
objects appear as faint smudges on SDSS images that are easily
confused with low surface brightness galaxies.

Under the assumption that the new GCs are all very old, their spread
in $(g-i)_{0}$ color suggests that they span a wide range in metal
abundance, which from Figure 1 of \citet{pea11} we roughly estimate as
$-2.5 < \textrm{[Fe/H]} < -0.2$.  The clusters with $R_{gc} < 30$ kpc are on
average redder and hence probably more metal-rich than the ones with
$R_{gc} > 30$ kpc.  We caution, however, that the reddening
values of the clusters are uncertain (see above).   

Figure 2 displays the r-band images of the objects that we classify as
clusters, which were cropped from the fits files that we downloaded
from the SDSS.  Cluster SDSS7 is the faint object that is touching on
the West the brighter object B320D, which is listed in the RBCv5 as a
galaxy.  Although we did not discover any clusters resembling the
extended clusters discussed above, Figure 2 shows that the new clusters
exhibit a wide range in brightness, central concentration, and degree
to which they are resolved into stars.  This last property may be due
to the range in distances from us, which for the clusters most remote
from M31 are expected to vary by $\sim100$ kpc around a mean of $\sim780$
kpc.

The spatial distribution of the new GCs is shown in
Figure 3.  There are six most remote GCs, with $R_{gc} > 100$
kpc, and the farthest one is at $R_{gc} = 158$ kpc, presenting a diffuse
nature with $R_{h} = 8.6$ and $M_{r} = -7.0$.  Two of these remote clusters
display a shorter projected distance to M33 than to M31.  One of these
clusters, SDSS13, lies $\sim6\farcm5$ from the And II dwarf spheroidal
galaxy, which is, however, sufficiently far that they may be
unrelated.

Table 2 lists the 75 candidate clusters, highlighting the 23 that we
judged with higher confidence.  Several of this last group display a
diffuse nature.  In Figure 4 we show a comparison between an HC
candidate cluster C35 of Table 2, and the extended cluster HEC11
\citep{hux08}, as an example of our classification criteria for
HC candidate clusters of diffuse type.  As Figure 4 shows, the r-band
SDSS images of C35 and HEC11 are similar in appearance, but there
is not an unambiguous sign that C35 is on the verge of resolving into
stars.  For this reason it not listed in Table 1 as a cluster.  We
list C35 and our other candidate clusters in Table 2 in order to
assist other observers who have the higher resolution data necessary
to make a definitive statement about their cluster status.

The newly discovered clusters represent only 0.02\% of the total
number of objects visually inspected in this survey.  If we combine
them with the 75 candidate clusters, even though the majority of these
candidates may not be clusters, we reach 0.09\% of the total, an
extremely small number of positive results among a vast population of
faint contaminant galaxies or ambiguous objects that could not be
resolved in the SDSS imaging.  In any case the results of our search
for GCs in more remote areas of M31 outer halo have
provided evidence for their existence up to $\sim160$ kpc of projected
galactocentric distance.  It is possible that other GCs reside in the
areas of our survey but were outside of our selection and
classification limits.  It is also possible that other GCs reside
beyond the edges of our regions.  A second more complete survey is
already underway and the results will be published in an upcoming
paper.

\acknowledgements

We gratefully acknowledge the technical support provided by Gabriele
Zinn throughout this project, which greatly facilitated its
completion.  This research has been supported by NSF grant AST-1108948
to Yale University.  This project would not have been possible without
the public release of the data from the Sloan Digital Sky Survey III
and the very useful tools that the SDSS has provided for accessing and
examining the publicly released data.  Funding for SDSS-III has been
provided by the Alfred P. Sloan Foundation, the Participating
Institutions, the National Science Foundation, and the U.S. Department
of Energy Office of Science. The SDSS-III Web site is
http://www.sdss3.org/.

SDSS-III is managed by the Astrophysical Research Consortium for the
Participating Institutions of the SDSS-III Collaboration including the
University of Arizona, the Brazilian Participation Group, Brookhaven
National Laboratory, University of Cambridge, Carnegie Mellon
University, University of Florida, the French Participation Group, the
German Participation Group, Harvard University, the Instituto de
Astrofisica de Canarias, the Michigan State/Notre Dame/JINA
Participation Group, Johns Hopkins University, Lawrence Berkeley
National Laboratory, Max Planck Institute for Astrophysics, Max Planck
Institute for Extraterrestrial Physics, New Mexico State University,
New York University, Ohio State University, Pennsylvania State
University, University of Portsmouth, Princeton University, the
Spanish Participation Group, University of Tokyo, University of Utah,
Vanderbilt University, University of Virginia, University of
Washington, and Yale University.

\clearpage
\begin{figure}
\epsscale{0.80}
\plotone{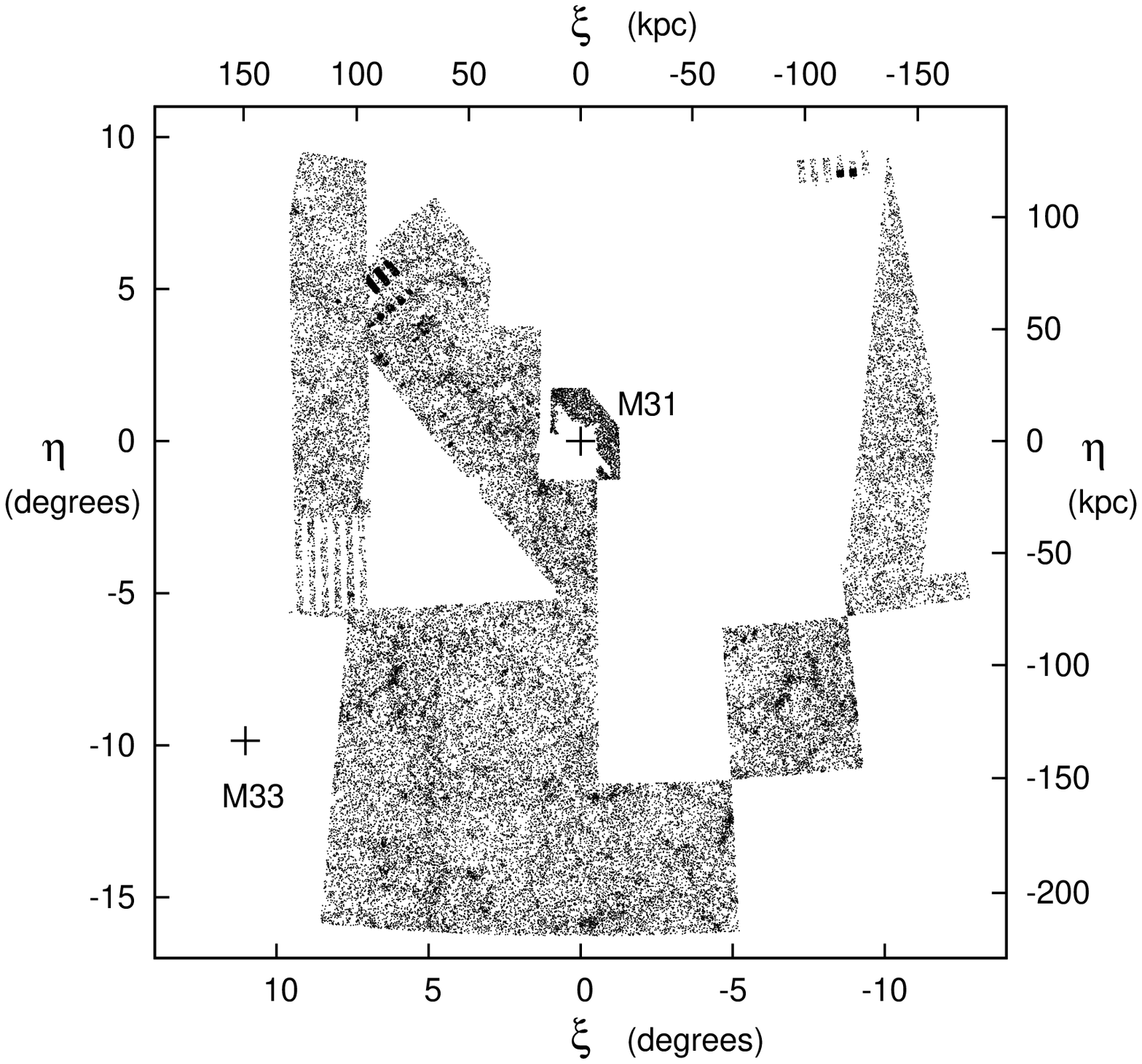}
\caption{Surface density map of the objects visually inspected in the survey regions around M31 and covered by SDSS.}
\end{figure}

\clearpage

\begin{figure}
\epsscale{0.50}
\plotone{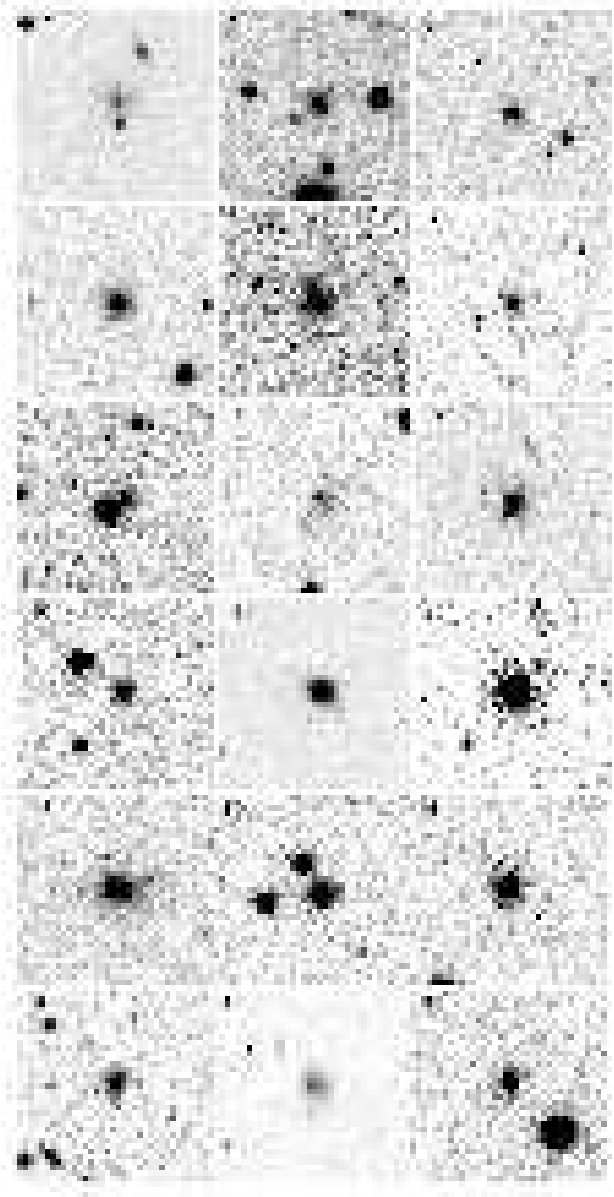}
\caption{r-band images ($90\arcsec \times 90\arcsec$, north at the top, east to the left) of the new clusters from SDSS images.  They are in R.A. order, according to their ID number.}
\end{figure}

\clearpage

\begin{figure}
\epsscale{0.80}
\plotone{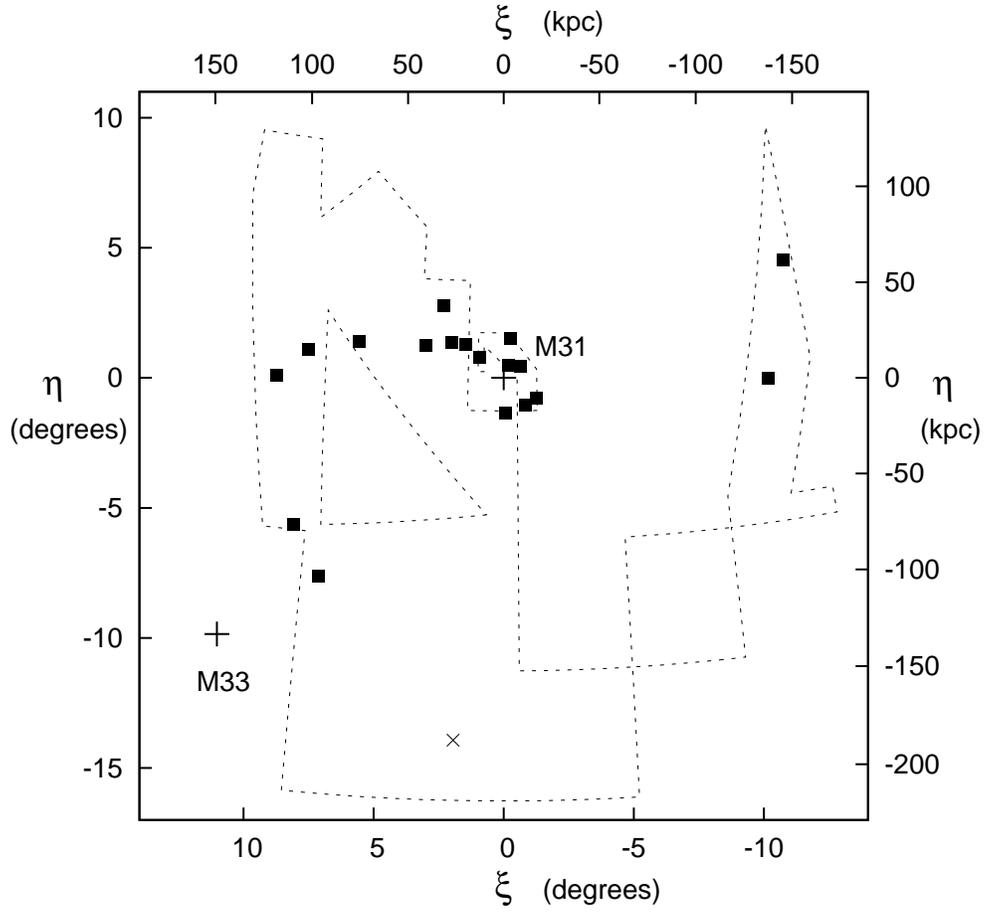}
\caption{Map of the spatial distribution of the newly discovered
  clusters at the distance of M31 (filled squares).  The $``\times''$
  represents the HC candidate $C35^{1d}$ (see the text).  The dashed line
  marks the footprint of SDSS that covered our survey regions}
\end{figure}

\clearpage

\begin{figure}
\epsscale{0.80}
\plotone{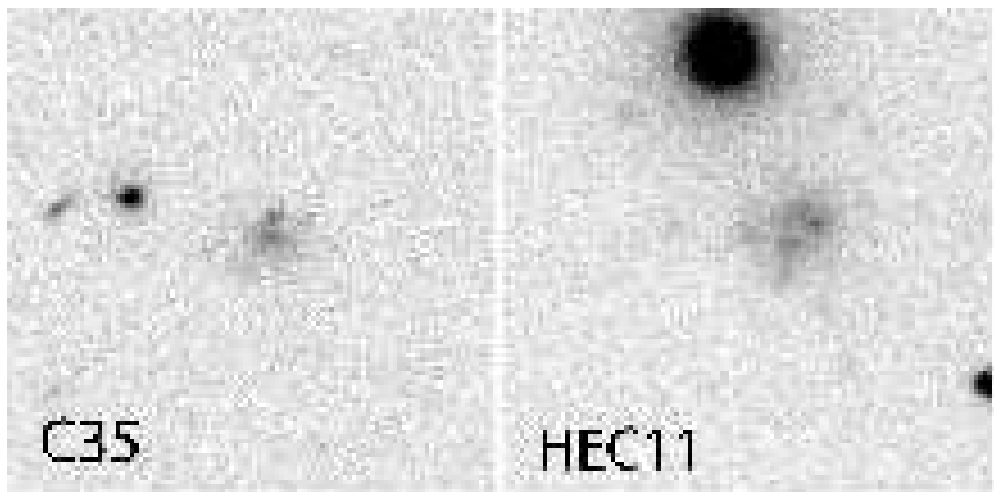}
\caption{Comparison between the HC candidate cluster $C35^{1d}$ and HEC11.  r-band images ($90\arcsec \times 90\arcsec$).}
\end{figure}




\begin{deluxetable}{cccccccc}




\tablecaption{Properties of the new globular clusters}

\tablenum{1}

\tablehead{\colhead{ID} & \colhead{RA(2000)} & \colhead{Dec(2000)} & \colhead{$r_{0}$} & \colhead{$(g-i)_{0}$} & \colhead{$R_{gc}$} & \colhead{$R_{h}$} & \colhead{$M_{r}$} \\ 
\colhead{} & \colhead{} & \colhead{} & \colhead{} & \colhead{} & \colhead{(kpc)} & \colhead{(pc)} & \colhead{} } 

\startdata
SDSS1\tablenotemark{d} & 00:36:01.8 & 40:29:50 & 17.35 & 1.14 & 20 & 10.5 & -7.1 \\
SDSS2 & 00:38:26.9 & 40:12:35 & 18.03 & 0.81 & 18 & 6.1: & -6.4 \\
SDSS3 & 00:39:13.1 & 41:42:08 & 18.19 & 1.02 & 11 & 5.9 & -6.3 \\
SDSS4 & 00:41:18.0 & 42:46:16 & 17.73 & 1.05 & 21 & 7.2 & -6.7 \\
SDSS5 & 00:41:47.2 & 41:44:10 & 17.50 & 1.01 & 7 & 10.5 & -7.0 \\
SDSS6 & 00:42:27.6 & 39:55:28 & 18.40 & 0.96 & 18 & 7.1 & -6.1 \\
SDSS7 & 00:47:41.1 & 42:04:17 & 19.02 & 1.07 & 17 & 5.8: & -5.4 \\
SDSS8\tablenotemark{d} & 00:50:36.3 & 42:31:50 & 18.17 & 0.48 & 26 & 9.1 & -6.3 \\
SDSS9 & 00:53:39.6 & 42:35:15 & 17.14 & 0.69 & 33 & 7.2 & -7.3 \\
SDSS10 & 00:55:28.1 & 43:59:31 & 18.62 & 0.58 & 49 & 5.1 & -5.8 \\
SDSS11 & 00:58:56.4 & 42:27:38 & 15.68 & 0.70 & 44 & 4.0 & -8.9 \\
SDSS12 & 01:12:47.0 & 42:25:25 & 16.85 & 0.68 & 78 & 5.0 & -7.6 \\
SDSS13 & 01:16:41.7 & 33:19:25 & 17.00 & 0.48 & 141 & 10.0 & -7.5 \\
SDSS14 & 01:22:20.7 & 35:11:35 & 17.79 & 0.92 & 134 & 8.2 & -6.7 \\
SDSS15 & 01:23:03.5 & 41:55:11 & 16.85 & 0.71 & 103 & 4.0 & -7.6 \\
SDSS16 & 01:29:02.2 & 40:47:09 & 18.17 & 0.80 & 119 & 6.6 & -6.3 \\
SDSS17\tablenotemark{d} & 23:41:50.0 & 44:50:07 & 17.46 & 0.44 & 158 & 8.6 & -7.0 \\
SDSS18 & 23:49:09.7 & 40:27:30 & 18.15 & 0.82 & 137 & 7.1 & -6.3 \\
\enddata


\tablecomments{Identification name used in this work, coordinates, Petrosian magnitude and color corrected for extinction, projected galactocentric distance, half-light radius, absolute magnitude at M31's distance.  \tablenotemark{d}Diffuse cluster.}


\end{deluxetable}




\begin{deluxetable}{cccc|cccc}




\tablecaption{Candidate globular clusters}

\tablenum{2}

\tablehead{\colhead{ID} & \colhead{RA(2000)} & \colhead{Dec(2000)} & \colhead{$r_{0}$} & \colhead{ID} & \colhead{RA(2000)} & \colhead{Dec(2000)} & \colhead{$r_{0}$} \\
\colhead{} & \colhead{} & \colhead{} & \colhead{} & \colhead{} & \colhead{} & \colhead{} & \colhead{} }
\
\startdata
C1\tablenotemark{1d} & 00:00:37.9 & 32:25:04 & 18.74 & C39 & 00:52:34.6 & 43:18:25 & 19.64 \\
C2 & 00:08:19.0 & 34:28:07 & 17.77 & C40 & 00:54:06.3 & 29:55:18 & 18.45 \\
C3 & 00:08:34.5 & 34:37:38 & 16.21 & C41 & 01:00:12.7 & 34:00:43 & 18.54 \\
C4 & 00:11:46.6 & 32:48:12 & 17.09 & C42 & 01:02:32.1 & 25:00:07 & 18.13 \\
C5 & 00:13:41.4 & 31:16:35 & 18.20 & C43 & 01:03:15.6 & 29:28:34 & 17.69 \\
C6 & 00:15:04.9 & 32:02:12 & 18.32 & C44 & 01:04:13.4 & 28:35:24 & 18.06 \\
C7\tablenotemark{1} & 00:23:07.6 & 27:30:47 & 17.58 & C45 & 01:05:43.0 & 30:56:43 & 17.35 \\
C8 & 00:26:35.6 & 26:58:31 & 17.31 & C46 & 01:06:06.6 & 27:45:30 & 18.75 \\
C9 & 00:32:16.5 & 25:16:56 & 18.04 & C47 & 01:06:14.8 & 34:01:15 & 17.73 \\
C10 & 00:33:59.7 & 28:42:30 & 17.93 & C48 & 01:06:40.5 & 32:29:59 & 18.05 \\
C11\tablenotemark{1} & 00:36:18.6 & 28:19:04 & 17.65 & C49 & 01:07:24.8 & 29:07:34 & 17.73 \\
C12 & 00:37:37.5 & 25:08:45 & 17.13 & C50\tablenotemark{1d} & 01:08:33.5 & 33:47:10 & 17.96 \\
C13 & 00:38:13.4 & 29:39:15 & 17.80 & C51 & 01:09:40.6 & 34:14:13 & 16.87 \\
C14\tablenotemark{1} & 00:39:32.3 & 40:51:17 & 17.85 & C52 & 01:10:50.8 & 44:44:38 & 17.38 \\
C15\tablenotemark{1d} & 00:40:09.5 & 39:55:30 & 17.34 & C53 & 01:13:22.3 & 25:22:37 & 17.56 \\
C16\tablenotemark{1d} & 00:40:14.0 & 39:02:33 & 18.67 & C54 & 01:13:54.4 & 28:46:06 & 18.11 \\
C17 & 00:40:31.9 & 38:11:12 & 18.11 & C55 & 01:14:29.6 & 46:06:07 & 17.07 \\
C18\tablenotemark{1} & 00:41:38.9 & 37:19:34 & 17.79 & C56\tablenotemark{1} & 01:17:36.0 & 46:11:18 & 18.42 \\
C19 & 00:42:05.4 & 29:35:02 & 18.81 & C57 & 01:18:09.8 & 29:14:09 & 18.01 \\
C20 & 00:42:09.2 & 38:56:15 & 17.81 & C58 & 01:19:43.8 & 33:09:20 & 18.24 \\
C21\tablenotemark{1d} & 00:42:54.5 & 29:39:34 & 18.61 & C59\tablenotemark{1} & 01:22:56.7 & 42:14:39 & 19.23 \\
C22\tablenotemark{1} & 00:43:03.7 & 32:08:37 & 18.56 & C60\tablenotemark{1d} & 01:26:10.8 & 43:49:11 & 18.43 \\
C23 & 00:43:32.0 & 33:10:04 & 17.33 & C61 & 01:27:37.6 & 38:07:05 & 16.64 \\
C24\tablenotemark{1} & 00:43:44.3 & 31:41:24 & 18.79 & C62\tablenotemark{1d} & 01:27:47.6 & 40:40:48 & 18.66 \\
C25\tablenotemark{1d} & 00:43:57.2 & 26:58:46 & 18.84 & C63\tablenotemark{1d} & 01:28:38.6 & 44:00:47 & 17.76 \\
C26\tablenotemark{1} & 00:44:01.0 & 30:42:01 & 17.72 & C64 & 01:30:27.4 & 48:45:15 & 18.41 \\
C27 & 00:45:40.2 & 37:47:11 & 17.95 & C65\tablenotemark{1} & 01:31:17.4 & 45:43:43 & 18.54 \\
C28 & 00:45:57.0 & 25:01:44 & 17.63 & C66 & 01:32:45.8 & 42:57:32 & 17.65 \\
C29 & 00:47:32.6 & 28:03:56 & 18.23 & C67 & 01:33:59.1 & 42:38:02 & 17.63 \\
C30 & 00:48:25.4 & 29:16:03 & 17.42 & C68 & 01:34:06.0 & 45:43:32 & 17.05 \\
C31\tablenotemark{1} & 00:49:33.1 & 34:52:00 & 18.76 & C69 & 01:34:39.5 & 44:05:41 & 18.35 \\
C32 & 00:49:37.5 & 33:44:54 & 16.79 & C70 & 01:36:14.1 & 45:37:35 & 17.21 \\
C33 & 00:50:22.5 & 41:51:35 & 17.38 & C71 & 23:40:21.5 & 46:00:22 & 18.53 \\
C34 & 00:51:12.0 & 43:33:35 & 17.54 & C72 & 23:41:09.4 & 47:54:19 & 17.25 \\
C35\tablenotemark{1d} & 00:51:24.2 & 27:19:04 & 18.80 & C73 & 23:44:01.5 & 38:57:20 & 18.86 \\
C36 & 00:51:32.6 & 41:57:24 & 17.11 & C74\tablenotemark{1} & 23:46:49.9 & 45:14:50 & 17.95 \\
C37 & 00:51:47.3 & 41:37:32 & 16.92 & C75\tablenotemark{1} & 23:48:40.9 & 39:37:45 & 17.79 \\
C38 & 00:51:59.0 & 27:30:20 & 18.71 \\

\enddata
\
\
\tablecomments{\tablenotemark{1}High confidence candidate\
\tablenotemark{d}Diffuse candidate}
\
\
\end{deluxetable}

\end{document}